\documentclass[12pt]{article}
\usepackage{amsmath}
\usepackage{amssymb}
\usepackage{amstext}
\usepackage{amsfonts}
\usepackage{graphicx,epsfig}
\usepackage{slashbox}
\usepackage{indentfirst}

\newtheorem{theorem}{Theorem}[section]
\newtheorem{proposition}[theorem]{Proposition}

\newcommand{\stl}{\scriptstyle}

\begin{document}

\baselineskip 20pt

\title{The Parametric Symmetry and Numbers of the Entangled Class of
$2 \times M \times N$ System}
\author{Xikun Li$^{a}$\;, Junli Li$^{a}$\;, Bin Liu$^{a}$\;,
Cong-Feng Qiao$^{a,b}$\footnote{corresponding author}\\[0.5cm]
{\small $a)$ Dept. of Physics, Graduate
University, the Chinese Academy of Sciences}  \\
{\small YuQuan Road 19A, 100049, Beijing, China}\\
{\small $b)$ Theoretical Physics Center for Science Facilities
(TPCSF), CAS}\\
{\small YuQuan Road 19B, 100049, Beijing, China}\\
}
\date{}
\maketitle

\begin{abstract}
We present in the work two intriguing results in the entanglement
classification of pure and true tripartite entangled state of
$2\times M\times N$ under stochastic local operation and classical
communication. (i) the internal symmetric properties of the nonlocal
parameters in the continuous entangled class; (ii) the analytic
expression for the total numbers of the true and pure entangled
class $2\times M \times N$ states. These properties help people to
know more of the nature of the $2\times M\times N$ entangled system.
\end{abstract}

\section{Introduction}

The understanding of entanglement is thought as at the heart of
Quantum Information Theory (QIT). Nowadays, apart from its
theoretical relevance in the testing of local realistic theories,
quantum entanglement has been shown to have more practical
applications, such as teleportation and super dense codings, etc
\cite{qcqi}. For this reason, the entanglement is regarded as the
key physical resource in QIT and draws intensive attention to its
qualitative and quantitative descriptions \cite{qu-entangle}.

The entanglement of the two qubit system now is thought to be well
understood \cite{Peres-ppt, 3H-condition, Entanglement-Formation}.
However things turn out to be more complicated in multipartite and
high dimensional systems. A distinguished feature of such systems is
that there exist different classes of entanglement. Two pure states
that can interrelated through stochastic local operations and
classical communication (SLOCC) are said to be in the same class of
entanglement which, on the experimental side, means that these two
states are able to carry out the same quantum informational tasks
with nonzero probabilities. Mathematically, two quantum states are
said to be SLOCC equivalent if they are connected by invertible
local operators (ILOs). Within this framework, it was found that
there exist two inequivalent ways for the entangled three-qubit pure
states \cite{three-qubit}. Though considerable effort has been
devoted to this subject, for the general states of multiqubit  only
up to four qubits are fully classified to the best of our knowledge
\cite{four-qubit,inductive-4}.

Among various multipartite quantum systems, the $2\times M\times N$
pure state system has been studied with many different methods. In
\cite{2MN-Chen}, Chen {\it et al} constructed the true entanglement
classes of $2\times M\times N$ which have finite entanglement
classes, and then they found the entanglement class with one
continuous parameter in the $2\times 4\times 4$ system using their
range criterion and Low-to-High Rank Generating Mode method
\cite{244-Chen}. There they also noticed that the continuous
parameter in the entanglement class is not totally free. Cornelio
and Piza proposed a different method based on the matrix
decompositions to classify the entanglement of such tripartite
systems \cite{Tripartite-qubit}, where only partial entanglement
classes are listed. In the previous works \cite{NN2} and \cite{MN2},
we have fully classified all the true tripartite entanglement class
of $2\times M\times N$ system using the matrix decomposition method.
Recently, Chitambar {\it et al} studied the classification of
$2\times M\times N$ using the elegant theory of matric pencils
\cite{matrix-pencil,c-matrix-pencil}.

In this paper, with the large amount of the enumerated entanglement
classes in \cite{NN2, MN2}, we investigate the nature of free
parameters of the continuous entanglement class in $2\times M\times
N$ system in detail. The content is arranged as follows. In section
2, we sort the parameters in the entanglement class into redundant
and nonlocal ones (nonlocal means that it cannot be eliminated by
ILOs), and show that for the nonlocal parameters there exist a
discrete symmetry within the same entanglement class. In section 3,
an analytic expression of the total number of true entanglement
classes of $2\times M\times N$ system is derived. Finally some
concluding remarks are given in section 4.

\section{The symmetry of the parameters}

\begin{figure}[t,m,u]
\centering
\scalebox{0.35}{\includegraphics{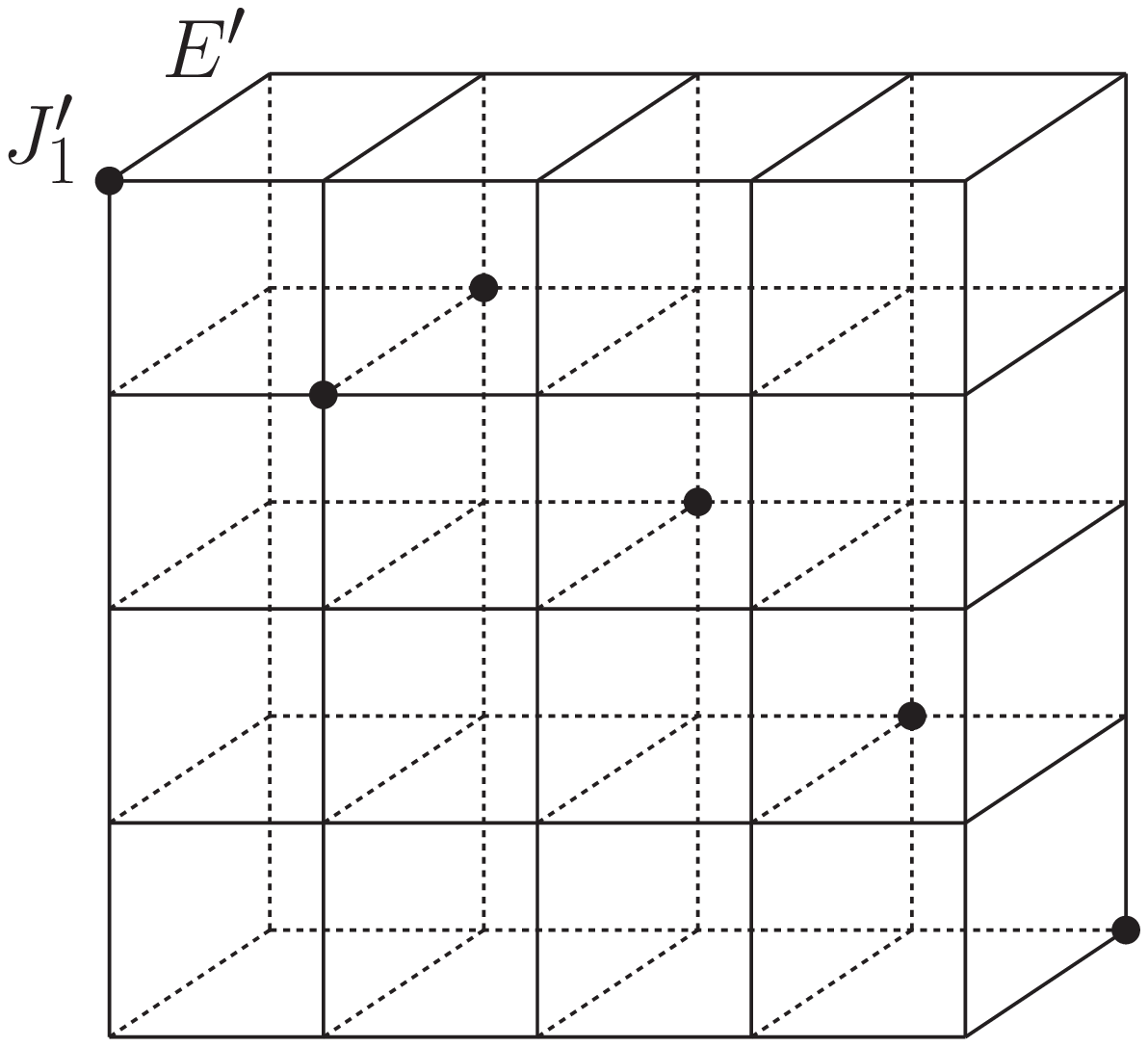}}%
\scalebox{0.35}{\includegraphics{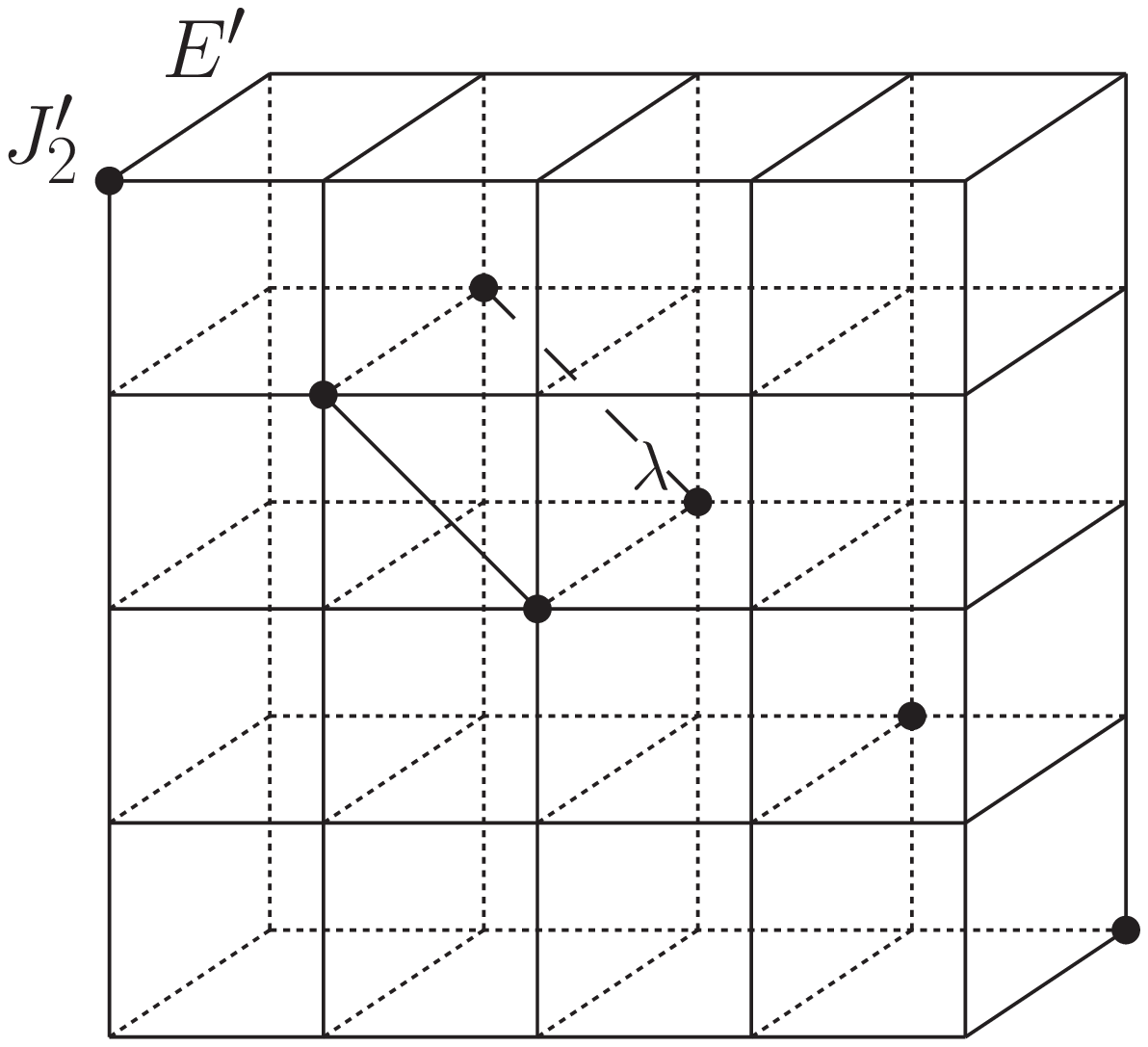}}%
\scalebox{0.35}{\includegraphics{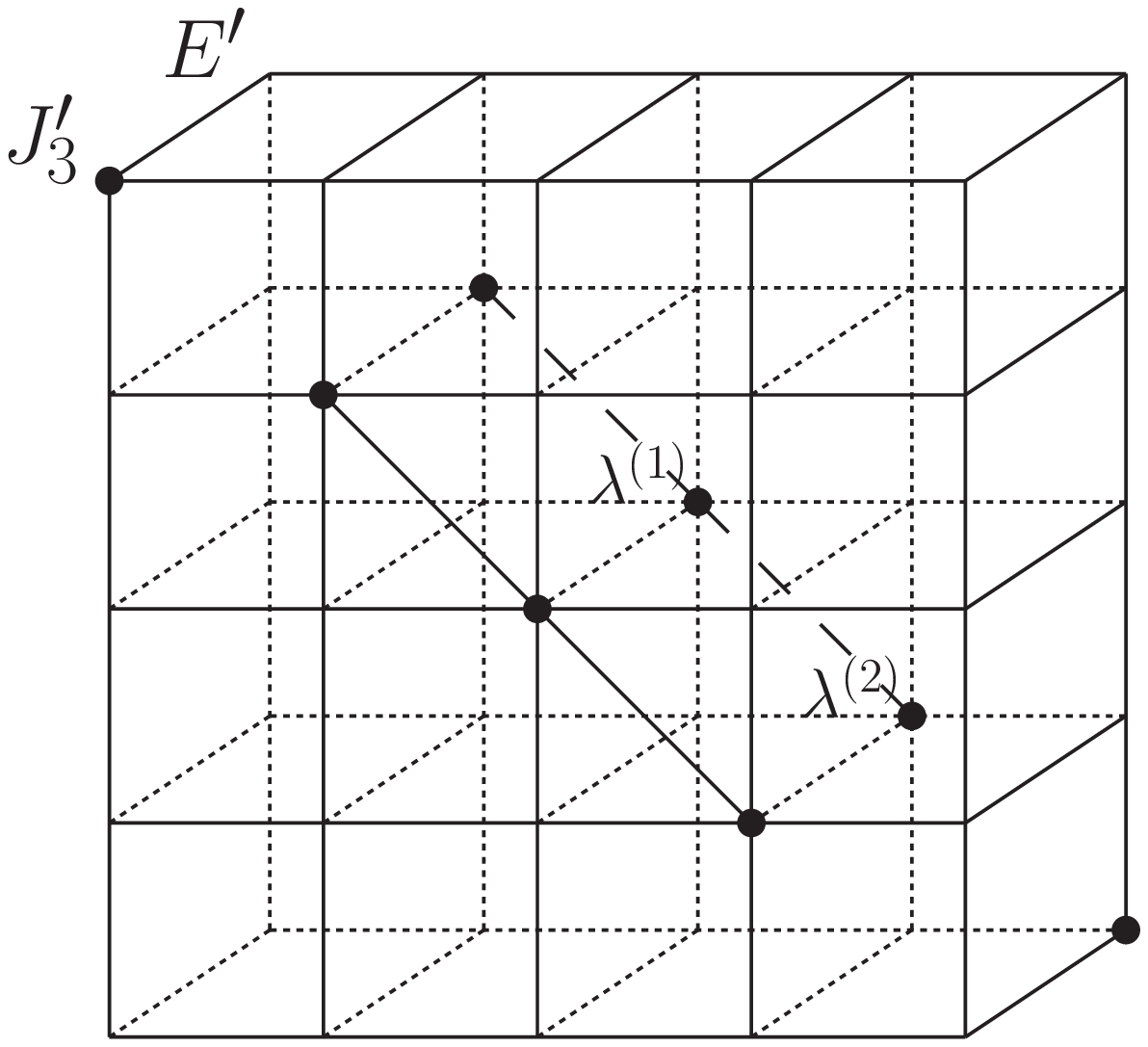}}%
\caption{\small The three cubic grids are the pictorial description
of $(E',J_i')$, where the solid nodes represent 1 if not specified
by $\lambda$, and the blank nodes are zeroes. Here
$\lambda,\lambda^{\scriptscriptstyle
(1)},\lambda^{\scriptscriptstyle (2)} \neq
\{0,1\}$.}\label{Pic-J-1-3}
\end{figure}

In the classification of the true tripartite entanglement states of
$2\times M\times N$ systems, lots of parameters were left in the
representative states, i.e., the eigenvalues of the Jordan forms
\cite{NN2,MN2}. Here we take three inequivalent entanglement classes
of $2\times 5\times 5$: $(E,J_1) \in c_{5,2}$, $(E,J_2) \in
c_{5,3}$, and $(E,J_3)\in c_{5,4}$ as examples. Because all of them
have the same $E$ which is a $5\times 5$ unit matrix, we only list
the $J_i$ to distinguish them
\begin{eqnarray}
J_1 = \begin{pmatrix}
\lambda_1 & 0 & 0 & 0 & 0 \\ 0 & \lambda_2 & 0 & 0 & 0 \\ 0 & 0 & 0 & 0 & 0 \\
0 & 0 & 0 & 0 & 0 \\ 0 & 0 & 0 & 0 & 0
\end{pmatrix}
J_2 = \begin{pmatrix}
\lambda_1 & 0 & 0 & 0 & 0 \\ 0 & \lambda_2 & 0 & 0 & 0 \\ 0 & 0 & \lambda_3 & 0 & 0 \\
0 & 0 & 0 & 0 & 0 \\ 0 & 0 & 0 & 0 & 0
\end{pmatrix}
J_3 = \begin{pmatrix}
\lambda_1 & 0 & 0 & 0 & 0 \\ 0 & \lambda_2 & 0 & 0 & 0 \\ 0 & 0 & \lambda_3 & 0 & 0 \\
0 & 0 & 0 & \lambda_4 & 0 \\ 0 & 0 & 0 & 0 & 0
\end{pmatrix} \; . \label{para-lam}
\end{eqnarray}
Here, $\forall i\neq j$, $\lambda_i\neq \lambda_j$, and $\forall i$,
$\lambda_{i} \neq 0$ . These parameters can be further simplified,
in the case of $J_1$, the following ILOs
\begin{eqnarray}
T & = & {\textstyle
\begin{pmatrix}
\frac{\lambda_2}{\lambda_1-\lambda_2} &
\frac{-\lambda_2}{\lambda_1(\lambda_1-\lambda_2)} \\ 0 &
\frac{1}{\lambda_1}
\end{pmatrix}}  \,, Q = E \; , \\
P & = & \text{diag}\{ {\textstyle 1, \frac{\lambda_1}{\lambda_2},
\frac{\lambda_1-\lambda_2}{\lambda_2} ,
\frac{\lambda_1-\lambda_2}{\lambda_2},
\frac{\lambda_1-\lambda_2}{\lambda_2}}\}   \;,
\end{eqnarray}
will make
\begin{eqnarray}
T \begin{pmatrix} PEQ \\ PJ_1Q
\end{pmatrix} = \begin{pmatrix} E' \\ J_1'
\end{pmatrix} \; , \label{TPQ-Action}
\end{eqnarray}
where $E'=\text{diag}\{ 0,1,1,1,1 \}$, $J_1'=\text{diag}\{ 1,1,0,0,0
\}$ (see Fig.(\ref{Pic-J-1-3})). Apparently there is no parameters
in $(E',J_1')$ now, so we call this kind of parameters in $(E,J_1)$
that can be factor out the entangled states the `redundant
parameters'. Similarly, the $(E,J_2)$ and $(E,J_3)$ can be
transformed into the form of $(E',J_2')$ and $(E',J_3')$ (see
Fig.(\ref{Pic-J-1-3})).

Consider a generally case of $2\times N\times N$ entangled state
\begin{eqnarray}
\begin{pmatrix}
E \\ J
\end{pmatrix} = \begin{pmatrix}
\text{diag}\{\;1\, ,\hspace{0.1cm} 1\, ,\cdots\,,\hspace{0.1cm}1\hspace{0.2cm},
\underbrace{1,\cdots,1}_{N-m}\} \\
\text{diag}\{\lambda_1,\lambda_2,\cdots
\,,\lambda_{m},\overbrace{0,\cdots,0}\}
\end{pmatrix} \; , \label{EJ-NN}
\end{eqnarray}
where $\lambda_i \in \mathbb{C}$; $\forall i\neq j$, $\lambda_i\neq
\lambda_j$; and $\lambda_{i} \neq 0$. With the following invertible
operators
\begin{eqnarray}
T & = &
\begin{pmatrix}
\frac{\lambda_2}{\lambda_1-\lambda_2} &
\frac{-\lambda_2}{\lambda_1(\lambda_1-\lambda_2)} \\ 0 &
\frac{1}{\lambda_1}
\end{pmatrix} \; , \; Q =E \; , \label{lambda-TQ} \\ P & = & \text{diag}\{ 1,\frac{\lambda_1}{\lambda_2},
\cdots, \frac{\lambda_1}{\lambda_{m}}, \frac{\lambda_1 -
\lambda_2}{\lambda_2},\cdots,\frac{\lambda_1 -
\lambda_2}{\lambda_2}\}\; , \label{lambda-P}
\end{eqnarray}
we have
\begin{eqnarray}
T\begin{pmatrix} PEQ \\ PJQ
\end{pmatrix} =
\begin{pmatrix}
E' \\ J'
\end{pmatrix} & = &
\begin{pmatrix}
\text{diag}\{0,1,\lambda^{(1)},\cdots,\lambda^{(m-2)},\underbrace{1,\cdots,1}_{N-m}\}
\\ \text{diag}\{1,1,\hspace{0.2cm}1\hspace{0.2cm}, \cdots,\;1\hspace{0.8cm},\overbrace{0,\cdots,0}\}
\end{pmatrix} \; , \label{loop-m}
\end{eqnarray}
where $\lambda^{(i)} =
\frac{(\lambda_1-\lambda_{i+2})}{(\lambda_1-\lambda_2)} \cdot
\frac{\lambda_2}{\lambda_{i+2}}$, and $\lambda^{(i)}\notin \{ 0,1
\}$. For $\lambda^{(i)}$s in Eq.(\ref{loop-m}), the following
proposition holds (see Appendix \ref{Appendix-1} for the proof)
\begin{proposition} \label{prop2.1}
The parameters $\lambda^{(i)}$s in the entanglement classes
\begin{eqnarray}
\begin{pmatrix}
E \\ J
\end{pmatrix} =
\begin{pmatrix}
\text{diag}\{0,1,\lambda^{(1)},\cdots,\lambda^{(m-2)},\underbrace{1,\cdots,1}_{N-m}\}
\\ \text{diag}\{1,1,\hspace{0.2cm}1\hspace{0.2cm}, \cdots,\;1\hspace{0.8cm},\overbrace{0,\cdots,0}\}
\end{pmatrix} \nonumber
\end{eqnarray}
are nonlocal parameters which can not be eliminated via ILO
transformations.
\end{proposition}
From this proposition we can infer that there are at most $N-3$
nonlocal parameters in $2\times N\times N$ entanglement classes.

Now all the parameters are sorted into two categories: one including
the redundant parameters, which can be eliminated out of the states
through ILOs ($\lambda$s in $(E,J_1)$); the other possesses nonlocal
properties (properties invariant under ILOs) which can not be
eliminated through the ILOs and will keep staying in the entangled
states as continuous parameters ($\lambda$s in $(E,J_2'),(E,J_3')$
). However, there exist residual symmetries on the nonlocal
parameters under ILOs. Take the entanglement class $(E',J_2')$ in
Fig.(\ref{Pic-J-1-3}) as an example, there exist the following
transformations
\begin{eqnarray}
\begin{pmatrix}
\text{diag}\{ 0,1,\lambda,1,1\}\\
\text{diag}\{1,1,1,0,0\}
\end{pmatrix} &\xrightarrow{F}&
\begin{pmatrix}
\text{diag}\{ 0,1,\frac{1}{\lambda},1,1\}\\
\text{diag}\{1,1,1,0,0\}
\end{pmatrix} \; ,\\
\begin{pmatrix}
\text{diag}\{ 0,1,\lambda,1,1\}\\
\text{diag}\{1,1,1,0,0\}
\end{pmatrix} &\xrightarrow{G}&
\begin{pmatrix}
\text{diag}\{ 0,1,1\!-\!\lambda,1,1\}\\
\text{diag}\{1,1,\hspace{0.3cm}1\hspace{0.3cm},0,0\}
\end{pmatrix} \; .
\end{eqnarray}
Here, transformation $G,F$ can be realized by the following ILOs:
\begin{eqnarray}
G &=& T\otimes P \otimes Q \nonumber \\ &=&
\begin{pmatrix}
-1 & 1 \\ 0 & 1
\end{pmatrix} \otimes
\begin{pmatrix}
0 & 1 & 0 & 0 & 0 \\ 1 & 0 & 0 & 0 & 0 \\ 0 & 0 & 1 & 0 & 0 \\ 0 & 0
& 0 & -1 & 0 \\ 0 & 0 & 0 & 0 & -1
\end{pmatrix} \otimes \begin{pmatrix}
0 & 1 & 0 & 0 & 0 \\ 1 & 0 & 0 & 0 & 0 \\ 0 & 0 & 1 & 0 & 0 \\ 0 & 0
& 0 & 1 & 0 \\ 0 & 0 & 0 & 0 & 1
\end{pmatrix} \; , \label{G-ILO}
\end{eqnarray}
\begin{eqnarray}
F &=& T\otimes P \otimes Q \nonumber \\ &=&
\begin{pmatrix}
\frac{1}{\lambda} & 0 \\ 0 & 1
\end{pmatrix} \otimes
\begin{pmatrix}
1 & 0 & 0 & 0 & 0 \\ 0 & 0 & 1 & 0 & 0 \\ 0 & 1 & 0 & 0 & 0 \\ 0 & 0
& 0 & \lambda & 0 \\ 0 & 0 & 0 & 0 & \lambda
\end{pmatrix} \otimes \begin{pmatrix}
1 & 0 & 0 & 0 & 0 \\ 0 & 0 & 1 & 0 & 0 \\ 0 & 1 & 0 & 0 & 0 \\ 0 & 0
& 0 & 1 & 0 \\ 0 & 0 & 0 & 0 & 1
\end{pmatrix} \; , \label{F-ILO}
\end{eqnarray}
where $T,P,Q$ act on the quantum state as in Eq.(\ref{TPQ-Action}).
The transformed entangled states under $F,G$ are SLOCC equivalent
with their initial states. If we assign
\begin{eqnarray}
F(\lambda) = \frac{1}{\lambda} \; , \; G(\lambda) = 1-\lambda \; ,
\nonumber \\  FG(\lambda) = F(G(\lambda)) = \frac{1}{1-\lambda} \; ,
\end{eqnarray}
then the $F,G$ operations generate a group, i.e., $\{E, F, G, GFG,
FG, GF \}$, which is isomorphic to $S_3$ group \cite{255class}.

Considering the general case of $(E',J')$ in Eq.(\ref{loop-m}), we
can represent the $m-2$ parameters in a row vector
\begin{eqnarray}
(\lambda^{(1)},\cdots,\lambda^{(m-2)}) \doteq
\begin{pmatrix}
\text{diag}\{ 0,1,\lambda^{(1)},\lambda^{(2)},\cdots, \lambda^{(m-2)},1, \cdots, 1\}\\
\text{diag}\{ 1,1,1\hspace{0.4cm}, 1\hspace{0.5cm},\cdots,
\hspace{0.3cm} 1\hspace{0.6cm},\, 0,\cdots, 0\}
\end{pmatrix} \; .
\end{eqnarray}
Here $\doteq$ means represented. Define
\begin{eqnarray}
A_{i}(\lambda^{(1)}, \cdots, \lambda^{(i)}, \lambda^{(i+1)}, \cdots,
\lambda^{(m-2)}) & = &
(\lambda^{(1)},\cdots, \lambda^{(i+1)}, \lambda^{(i)},\cdots, \lambda^{(m-2)}) \; ,\\
F(\lambda^{(1)}, \cdots,\lambda^{(m-2)})
& = & (\frac{\lambda^{(1)}}{\lambda^{(m-2)}}, \frac{\lambda^{(2)}}{\lambda^{(m-2)}},
\cdots, \frac{1}{\lambda^{(m-2)}}) \; , \\
G(\lambda^{(1)}, \cdots,\lambda^{(m-2)}) & = & (1-\lambda^{(1)},
\cdots, 1-\lambda^{(m-2)}) \; ,
\end{eqnarray}
where all the transformation $A,F,G$ can be realized as that of
Eqs.(\ref{F-ILO},\ref{G-ILO}). If we assign the operators $A_i =
\sigma_i, F = \sigma_{m-2}, G = \sigma_{m-1}$, it can be verified
that
\begin{equation}
\left\{ \begin{aligned}
\sigma_{i}^2 & = 1 \; \\
\sigma_i\sigma_j &= \sigma_j\sigma_i  \hspace{1cm}\text{if $|j- i|
\geq 2$}\; \\ \sigma_i\sigma_{i+1} \sigma_{i} &=
\sigma_{i+1}\sigma_{i} \sigma_{i+1}
\end{aligned} \right. \; .
\end{equation}
These are the generators of the $S_{m}$ symmetric group. If the
dimension $N=m+1$, then there is another additional symmetry
operation $H = \sigma_{m} $ where
\begin{eqnarray}
H(\lambda^{(1)}, \cdots,\lambda^{(m-2)}) = (\frac{1}{\lambda^{(1)}},
\cdots, \frac{1}{\lambda^{(m-2)}}) \; , \label{n=m+1}
\end{eqnarray}
then$(A_i,F,G,H)$ will generate  an $S_{m+1}$ group.

\section{The total number of the entanglement classes}

Regard the entanglement class with nonlocal parameters in a
representative state (e.g., state $(E',J_2')$ in
Fig.(\ref{Pic-J-1-3})) as one continuous class, we have shown that
there are 61 classes in $2\times 6\times 7$ states \cite{MN2}. In
our classification schemes, the number of the entanglement classes
of sets $c_{N,l}$ in \cite{NN2} (or $c_{M,l}$ in \cite{MN2}) can be
counted by the number of Jordan forms, which are characterized by
Segre symbols. There is one case that do not correspond to the true
tripartite entanglement in $c_{N,l}$ with $(E,J)$ where $J=[
(\underbrace{11\cdots11}_{N}) ]$. This corresponds to the following
case
\begin{eqnarray}
\begin{pmatrix}
E \\ J
\end{pmatrix} =
\begin{pmatrix}
\text{diag}\{ 1,1,\cdots,1,1 \} \\ \text{diag}\{ 1,1,\cdots,1,1 \}
\end{pmatrix} \; , \label{eq-deltamn}
\end{eqnarray}
which is actually a bipartite $N\times N$ entangled state. It is
known that the generating function of the number of Segre symbols
$S(n)$ for $n\times n$ matrix is \cite{OEIS}
\begin{eqnarray}
\prod_{i=1}^{\infty} \frac{1}{(1-x^{i})^{P(i)}} = \sum_{n} S(n)x^{n}
\; ,
\end{eqnarray}
where $P(i)$ is the number of partitions of integer $i$.

\begin{figure}[t,m,u]
\centering
\scalebox{0.4}{\includegraphics{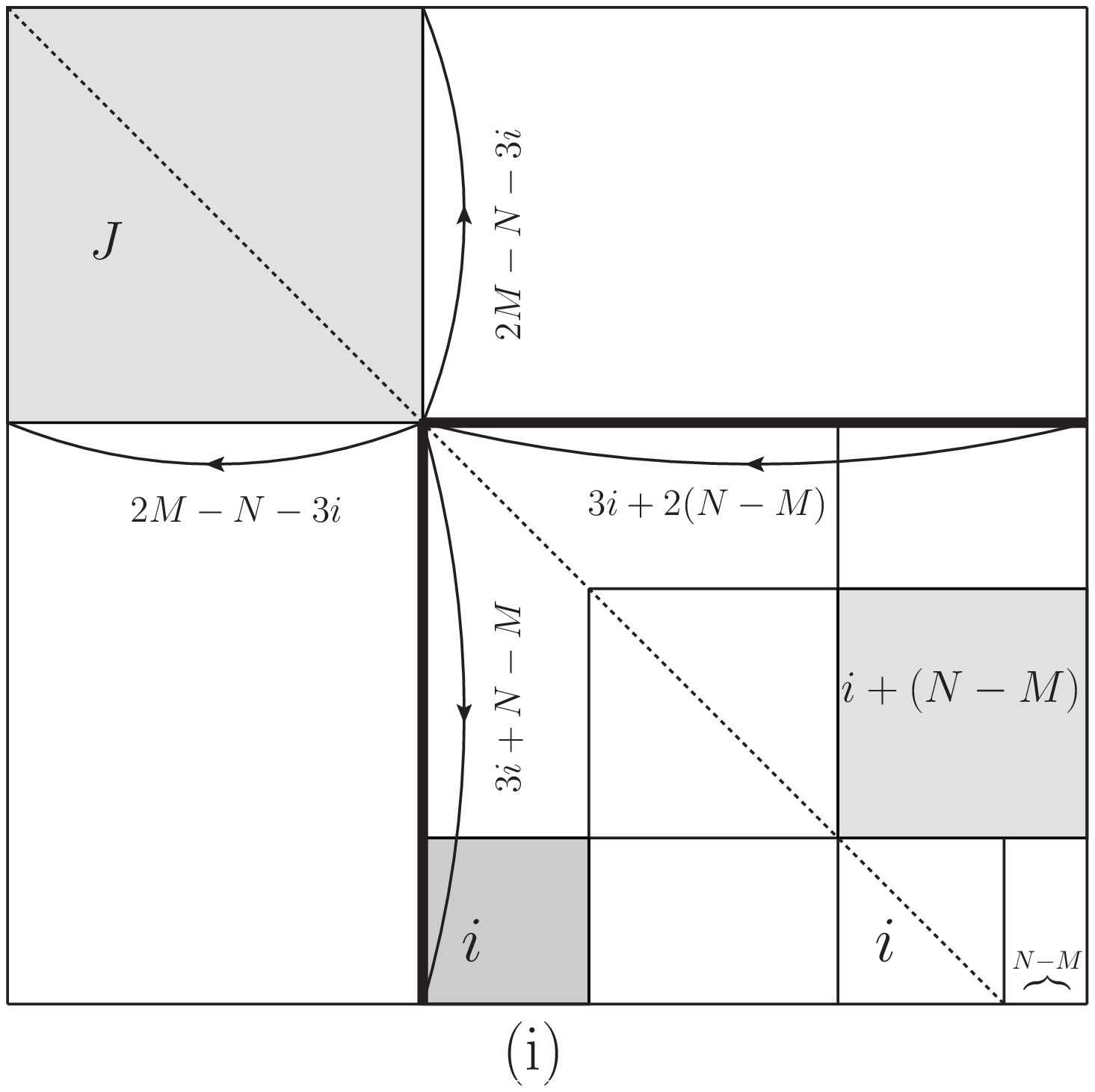}}%
\scalebox{0.4}{\includegraphics{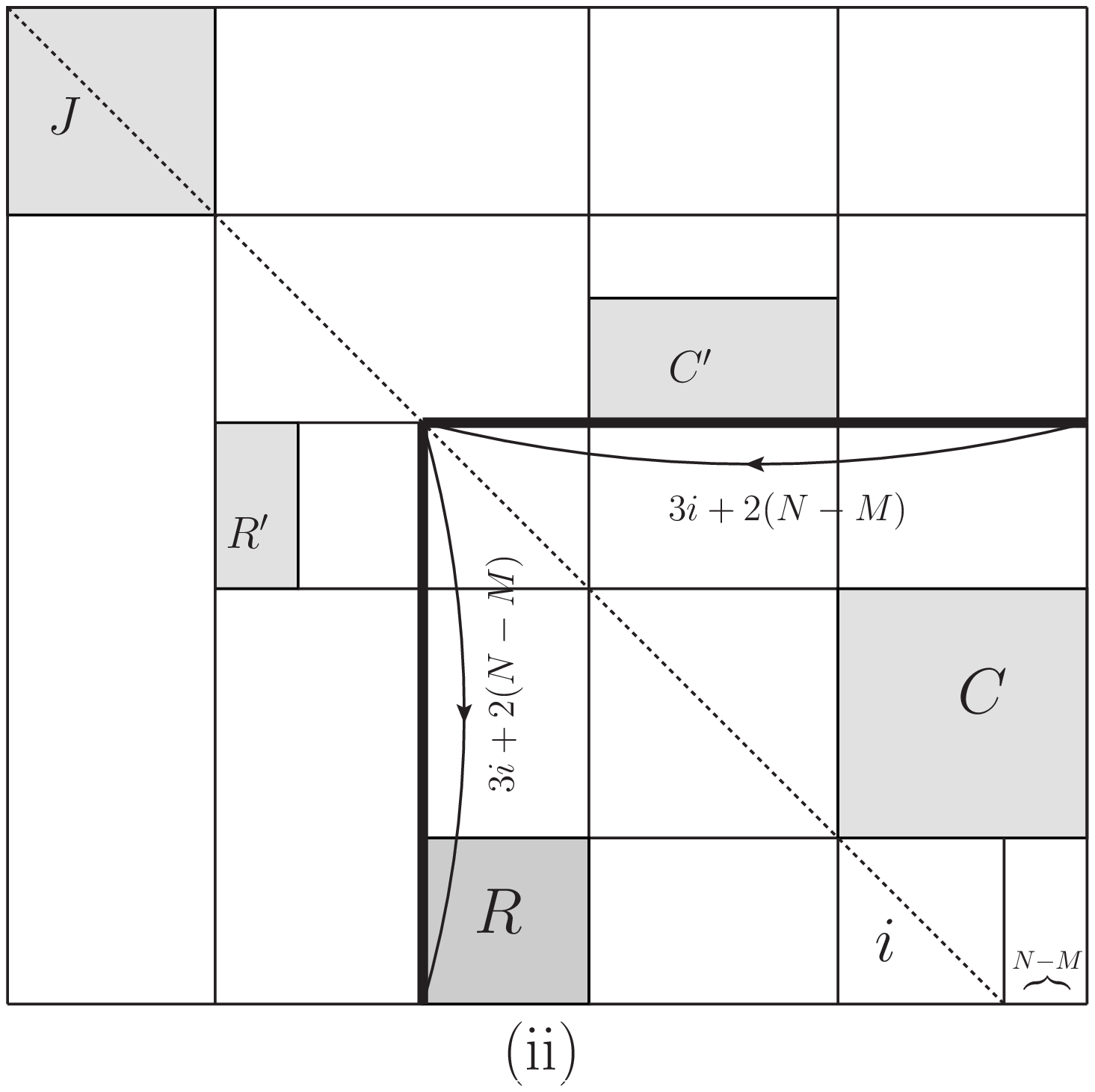}}
\caption{\footnotesize Two different forms of $\Gamma_{2}$ in
Eq.(\ref{Gamma_2}). The thick lines outlines the $B$ matrix. (i) the
initial $B$ matrix with the dimensions being labeled. (ii) In the
next step of enlargement of $B$ matrix, the rank of submatrices $C'$
and $R'$ must be less than that of $R$ and $C$.} \label{MatrixMN2}
\end{figure}

Consider the general entanglement sets $c_{M-i,l}$ of $2\times
M\times N$ system. The
canonical form of the matrix pair $\left(\begin{smallmatrix} \Gamma_{\!1} \\
\Gamma_{\!2} \end{smallmatrix}\right) \in c_{M-i,l}$ has the
following forms (see Eq.(49) of Ref.(\cite{MN2}))
\begin{eqnarray}
\Gamma_{\!1} = \left(
  \begin{array}{lll}
    E_{\scriptstyle (M-i)\times (M-i) } & {\bf 0} & {\bf 0} \\
    {\bf 0} & {\bf 0}_{\scriptstyle i\times i} & {\bf 0}_{\scriptstyle i\times (N-M)} \\
  \end{array}
\right) \; ,
\end{eqnarray}
and
\begin{eqnarray}
\Gamma_{\!2} = \left(\begin{array}{ll}
J_{d_J} & {\bf 0} \\
{\bf 0} & B_{(M-d_J) \times (N-d_J)}
\end{array}\right)\; , \label{Gamma_2}
\end{eqnarray}
where $d_J$ is the dimension of $J$ and ${\bf 0}$ are the zero
submatrices, see Fig.(\ref{MatrixMN2}). We can formally write the
number of inequivalent classes of the sets $\{c_{M-i,l}\}$ by
$\omega_{M,N}$ as follows
\begin{eqnarray}
\omega_{M,N}(i, d_J) = S(d_J)\cdot F_{r_B}\; .
\end{eqnarray}
Here $r_B$ represents the rank of $B$, $F_{r_B}$ is the number of
different forms of $B$ in $c_{M-i,l}$. The value of $F_{r_B}$ can be
deduced from the construction procedures of $B$, see
Fig.(\ref{MatrixMN2}). If we know $B_i$ (see the submatrix outlined
by the thick lines in (ii) of Fig.(\ref{MatrixMN2})), the
$B^{(i+1)}$ then can be constructed based on the rank of $R,C$ of
$B^{(i)}$. And the rank of $R',C'$ in $B^{(i+1)}$ must be less than
or equal to that of $R,C$ separately (see (ii) of
Fig.(\ref{MatrixMN2})). There will be three cases: (1), $r(C')=0$,
and all the $B$ matrices after $B^{(i+1)}$ will have
$r(C'')=0,\cdots$; (2), $r(R')=0$ which is the similar to (1); (3),
$r(C')\neq 0$ and $r(R')\neq 0$ which we can construct $B^{(i+2)}$
recursively. Translate this into mathematics, we can get the
following recursive formula of the number of $F_{r_B}$
\begin{eqnarray}
F_{r_B} = F(j,r,c) = F(j,r,0) + F(j,0,c) +
\sum_{m=1}^{r}\sum_{n=1}^{c}F(j-m-n,m,n) \; ,
\end{eqnarray}
where $r, c$ are the rank of $R,C$ associated with the corresponding
$B$ submatrix and the initial values are $r=i, c=i+N-M$ separately;
$ j=r_B- i- (i+N-M)$; $F(j,r,0)=f_{j}^{(r)}, F(j,0,c)=f_{j}^{(c)}$.
Here $f_{n}^{(m)}$ is the number of partitions of $n$ where the
maximum part is $m$ whose generating function is
\begin{eqnarray}
\prod_{k=1}^{m}\frac{1}{1-x^k} = \sum_{n}f_{n}^{(m)} x^{n} \; .
\end{eqnarray}
It can be verified that $F(0,r,c) = 1$, and we assume $F(-j,r,c) =
0$.

If the the initial matrix is $B_{(3i+N-M)\times (3i+2(N-M))}$ whose
rank is $(2i+N-M)$, see (i) of Fig.(\ref{MatrixMN2}), then there has
only one form. Thus there are $S(2M-N-3i)$ inequivalent classes,
\begin{eqnarray}
\omega_{M,N}(i, j=0)& = & S(2M-N-3i)\cdot F(j,R,C) \nonumber \\
& = & S(2M-N-3i)\cdot F(0,i,i+(N-M)) \nonumber \\
& = & S(2M-N-3i)\cdot 1\; .
\end{eqnarray}
If the rank of $B$ enlarged to $j+(2i+N-M)$, we have the number of
inequivalent classes $\omega_{M,N}(i,j)$ to be
\begin{eqnarray}
\omega_{M,N}(i,j)  = S(2M-N-3i-j)\cdot F(j,i,i+(N-M)) \; .
\end{eqnarray}
Specifically when $M=N$ and $i=j=0$, there is one case that does not
correspond to true entanglement (see Eq.(\ref{eq-deltamn})). Thus
the total number of inequivalent entanglement classes $\Omega(M,N)$
can be derived
\begin{eqnarray}
\Omega_{M,N} = \sum_{i=0}^{\stl \lfloor\frac{2M-N}{3}\rfloor}\,
\sum_{j=0}^{\stl (2M-N-3i)} \omega_{M,N}(i,j) -\delta_{MN} \; .
\label{Omega}
\end{eqnarray}
The above equation can be evaluated by computers for arbitrary given
$M$ and $N$. The values of $\Omega_{M,N}$ up to $2\times 10\times
10$ are listed in Table (\ref{Tb-1010}).

\begin{table}[htp]
\centering
\begin{tabular}{|l||l|l|l|l|l|l|l|l|l|l|}
\hline \backslashbox{$M$}{$N$} & 2 & 3 & 4 & 5 & 6 & 7 & 8 & 9 & 10 \\
\hline\hline 2 & 2 & 2 & 1 & 1 & 1 & 1 & 1 & 1 & 1 \\ \hline 3 & 2 &
6 & 5 & 2 & 1 & 1 & 1 & 1 & 1 \\ \hline 4 & 1 & 5 & 16 & 12 & 6 & 2
& 1 & 1 & 1 \\ \hline 5 & 1 & 2 & 12 & 34 & 28 & 14 & 6 & 2 & 1 \\
\hline 6 & 1 & 1 & 6 & 28 & 77 & 61 & 34 & 15 & 6 \\ \hline 7 & 1 &
1 & 2 & 14 & 61 & 157 & 133 & 74 & 36 \\ \hline 8 & 1 & 1 & 1 & 6 &
34 & 133 & 328 & 277 & 165 \\ \hline 9 & 1 & 1 & 1 & 2 & 15 & 74 &
277 & 655 & 572 \\ \hline 10 & 1 & 1 & 1 & 1 & 6 & 36 & 165 & 572 &
1309 \\ \hline
\end{tabular}
\caption{\small  The number of inequivalent classes of true
tripartite entanglement system of $2\times M\times N$ up to $10$.}
\label{Tb-1010}
\end{table}

From the table, it can be confirmed that the number of entanglement
classes of $2\times N\times N$ system increases exponentially with
the dimensions of system, i.e., $\Omega_{N,N} \sim 2^{N}$.

\section{Conclusions}

In this paper we have investigated two interesting features of the
entanglement classes of $2\times M\times N$ system. The continuous
entanglement classes with more than one nonlocal parameters come
into existence, and there exits a upper limit for the number of
nonlocal parameters. Meanwhile, there are some residual discrete
symmetries that remain exist under continuous ILO transformations of
$SL(2,\mathbb{C}), SL(M,\mathbb{C}), SL(N,\mathbb{C}),$ which are
isomorphic to symmetry groups. We also get an analytic expression
for the total number of inequivalent entanglement classes where the
same structured entanglement class with continuous parameters are
regard as the same class, and it indicates that the entanglement
classes are generally exponentially increasing with the dimensions
in $2\times N\times N$ system. With these results, the full
understanding of the entanglement classes of $2\times M\times N$
thus becomes promising. It is worth mentioning that the
classification of $2\times M\times N$ may shed some light on the
classification of other entangled systems, e.g. the entangled
$(2N+1)$-qubit system.

\newpage

\appendix{\bf\Large Appendix}

\section{Proof of Proposition \ref{prop2.1}} \label{Appendix-1}

\noindent {\bf Proof}:

Consider the following entangled class in Eq.(\ref{EJ-NN})(with
$N-m>1$)
\begin{eqnarray}
\begin{pmatrix}
E \\ J
\end{pmatrix} =
\begin{pmatrix}
\text{diag}\{\;1\, ,\hspace{0.1cm} 1\,
,\cdots\,,\hspace{0.1cm}1\hspace{0.2cm},
\underbrace{1,\cdots,1}_{N-m}\} \\
\text{diag}\{\lambda_1,\lambda_2,\cdots
\,,\lambda_{m},\overbrace{0,\cdots,0}\}
\end{pmatrix} \in c_{N,m} \; ,\nonumber
\end{eqnarray}
the ILO transformations $T,P,Q$ that apply on this class would
transform it into other forms, i.e., $(E,J')$. But standard form of
$(E,J')$ must be also in the set $c_{N,m}$, so we have
\begin{eqnarray}
\begin{pmatrix}
E \\ J'
\end{pmatrix} =
\begin{pmatrix}
\text{diag}\{\;1\, ,\hspace{0.1cm} 1\,
,\cdots\,,\hspace{0.1cm}1\hspace{0.2cm},
\underbrace{1,\cdots,1}_{N-m}\} \\
\text{diag}\{\lambda'_1,\lambda'_2,\cdots
\,,\lambda'_{m},\overbrace{0,\cdots,0}\}
\end{pmatrix} \in c_{N,m} \; .\nonumber
\end{eqnarray}
It is easy to verify that the operation $T,P,Q$ which makes
\begin{eqnarray}
(E,J) \xrightarrow{T,P,Q} (E,J')\; , \nonumber
\end{eqnarray}
will leads to $\lambda' = \frac{ t_{22} \lambda }{ t_{11} +
t_{12}\lambda}$, here $t_{ij}$ are matrix elements of $T$ see
Eq.(38) in \cite{NN2}. Here we neglect the subscripts of $\lambda$s
and $\lambda'$s, for there can be a change of the orders of
different $\lambda'$s induced by $P,Q$ operations. We conclude that
the ILO transformations ($T,P,Q$) induce a special linear fraction
transformations which keep $0$ invariant (i.e., $\lambda' = \frac{
t_{22} \lambda }{ t_{11} + t_{12}\lambda}$) on the eigenvalues of
$J$ in the entangled class $(E,J)$.

We apply the different ILO transformations introduced in
Eqs.(\ref{lambda-TQ},\ref{lambda-P}), the entangled state ($(E,J)
\in c_{N,m}$) is thus transformed into the form of
Eq.(\ref{loop-m}). In this form, there are $m-2$ $(m\geq 3)$
parameters $\lambda^{(k)}$, where $\lambda^{(k)} =
\frac{0-\lambda_2}{0-\lambda_{k+2}} \cdot
\frac{(\lambda_1-\lambda_{k+2})}{(\lambda_1-\lambda_2)} $ is the
cross ratio of $(0,\lambda_1,\lambda_2, \lambda_{k+2})$. Combined
with the argument in the previous paragraph, we have
\begin{proposition}\label{propA-1}
The cross ratio $\lambda^{(k)}$ is invariant under ILOs.
\end{proposition}

Now we proceed to prove Proposition \ref{prop2.1}. Suppose the $m-2$
$\lambda^{(i)}$ can be further transformed into a form with $m-2-l$
parameters $\lambda'^{(i)}$ via ILOs, where $l\geq 1$, then we would
have the following $m-2-l$ equations
\begin{eqnarray}
\lambda'^{(1)} & = & \lambda'^{(1)}(\lambda^{(1)},\cdots,
\lambda^{(m-2)}) \; ,\nonumber \\ & \cdots & \nonumber \\
\lambda'^{(m-2-l)} & = & \lambda'^{(m-2-l)}(\lambda^{(1)},\cdots,
\lambda^{(m-2)}) \; .\nonumber
\end{eqnarray}
Clearly there are less equations (there are $m-2-l$) than parameters
$\lambda^{(i)}$ (there are $m-2$). At least, there exists a
parameter, suppose $\lambda^{(k)}$, that can not be determined by
the $m-2-l$ equations by $\lambda'^{(i)}$s. This is equivalent to
say that the entangled states with continuous parameters
$\lambda^{(k)}$ are all equivalent to the same entangled state
specified by $\lambda'^{(i)}$s, therefore different value of
$\lambda^{(k)}$ are themselves ILO equivalent which contradicts
Proposition \ref{propA-1}. Thus we have the $m-2$ $\lambda^{(i)}$ in
Eq.(\ref{loop-m}) are nonlocal parameters which can not be further
eliminated by ILOs.

For the case of $N-m=1$, the proof is similar except inducing an
additional symmetry in Eq.(\ref{n=m+1}).

\vspace{.7cm} {\bf Acknowledgments} \vspace{.3cm}

This work was supported in part by the National Natural Science
Foundation of China(NSFC) under the grants 10935012, 10928510,
10821063 and 10775179, by the CAS Key Projects KJCX2-yw-N29 and
H92A0200S2, and by the Scientific Research Fund of GUCAS.



\begin{thebibliography}{99}

\bibitem{qcqi} M.A. Nielsen and I.L. Chuang,\
{\it Quantum Computation and Quantum Information}\ (Cambridge
University Press, Cambridge, England, 2000).

\bibitem{qu-entangle} Ryszard Horodecki, Pawe{\l} Horodecki, Micha{\l}
Horodecki, and Karol Horodecki, Rev. Mod. Phys. {\bf 81}, 865
(2009).

\bibitem{Peres-ppt} Asher Peres, Phys. Rev. Lett. {\bf 77}, 1413
(1996).

\bibitem{3H-condition} Micha{\l} Horodecki, Pawe{\l} Horodecki, and Ryszard
Horodecki, Phys. Lett. A {\bf 223}, 1 (1996).

\bibitem{Entanglement-Formation} William K. Wootters, Phys. Rev. Lett. {\bf
80}, 2245 (1998).

\bibitem{three-qubit} W. D\"{u}r, G. Vidal, and J.I. Cirac,\ Phys.\
Rev.\ A {\bf 62},\ 062314\ (2000).

\bibitem{four-qubit} F. Verstraete, J. Dehaene, B. De Moor, and H.
Verschelde,\ Phys.\ Rev.\ A {\bf 65},\ 052112\ (2002).

\bibitem{inductive-4} L. Lamata, J. Le\'{o}n, D. Salgado, and E.
Solano,\ Phys.\ Rev.\ A {\bf 75},\ 022318\ (2007).

\bibitem{2MN-Chen} Lin Chen and Yi-Xin Chen, Phys. Rev. A {\bf 73}, 052310
(2006).

\bibitem{244-Chen} Lin Chen, Yi-Xin Chen, and Yu-Xue Mei, Phys. Rev.
A {\bf 74}, 052331 (2006).

\bibitem{Tripartite-qubit} Marcio F. Cornelio and A. F. R. de Toledo Piza, Phys. Rev. A {\bf 73}, 032314
(2006).

\bibitem{NN2} Shuo Cheng, Junli Li, and Cong-Feng Qiao, J. Phys. A: Math. Theor. {\bf 43}, 055303
(2010).

\bibitem{MN2} Junli Li and Cong-Feng Qiao,
arXiv: 1001.0078(2010).

\bibitem{matrix-pencil} Eric Chitambar, Carl A. Miller, and Yaoyun
Shi, arXiv: 0911.1803.

\bibitem{c-matrix-pencil} Eric Chitambar, Carl A. Miller, and Yaoyun
Shi, arXiv: 0911.4058.

\bibitem{255class} Shuo Cheng, Junli Li, and Cong-Feng Qiao, Journal of the
Graduate School of the Chinese Academy of Sciences {\bf 3}, 303
(2009) (in chinese).

\bibitem{OEIS} N. J. A. Sloane, The On-Line Encyclopedia of Integer Sequences,
published electronically at
www.research.att.com/\~\,\!njas/sequences/, (2008).


\end{thebibliography}
\end{document}